\documentclass[preprint,aps,amsmath,amssymb,a4paper,showkeys,superscriptaddress]
	{revtex4}
\usepackage{graphicx}	
\usepackage{dcolumn}	
\usepackage{bm}		
\usepackage{epsfig}
\usepackage{comment}
\usepackage[usenames]{color}
\usepackage{bbm}	

\newcommand{\ket}[1]{| #1 \rangle}						  %
\newcommand{\overlap}[3]{\left\langle #1 \left| #2 \right| #3 \right\rangle}	  %
\newcommand{\etal}{{\it et al\/}\ }                                               %
\def \kp  {$\mathbf{k} \cdot \mathbf{p}$ }					  %
\def \p   {{\bf p}}								  %

\long\def\readword#1 {\strut\putmark#1 \next\relax}
  \def\margmark{$| \hskip10pt$}
  \def\putmark{%
      \vadjust{\vbox to 0pt{\vss\noindent
                            \llap{\strut\margmark}\vskip0pt}}%
      }
  \def\marginmarkson{\global\let\next =\readword\next}
  \def\marginmarksoff{\global\let\next=\relax}

\begin{document}

\title{Cooling of radiative quantum-dot excitons by terahertz radiation: A spin-resolved Monte Carlo carrier dynamics model}             		  %
\author{Fredrik Boxberg}	\email{fredrik.boxberg@tkk.fi}	  		  \affiliation{	  								  Laboratory of Computational Engineering, Helsinki University of Technology, P.O. Box 9203, FIN-02015 HUT, Finland		  }				  					  	  
\author{Jukka Tulkki}		
\affiliation{	  								  Laboratory of Computational~Engineering, Helsinki University of Technology, P.O. Box 9203, FIN-02015 HUT, Finland		  }				  					  	  
\author{Go Yusa}			
\affiliation{	  								  Institute of Industrial Science, The University of Tokyo, 4-6-1 Komaba, Meguro-Ku, Tokyo 153-8505, Japan						  }				  					  	  %
\author{Hiroyuki~Sakaki}	
\affiliation{
Institute of Industrial Science, The University of Tokyo, 4-6-1 Komaba, Meguro-Ku, Tokyo 153-8505, Japan						  }				  					  	  %
\date{\today}		  					 	 	  %
\begin{abstract}		  					  	  %
We have developed a theoretical model to analyze the
anomalous cooling of radiative quantum dot (QD) excitons by
THz radiation reported by Yusa \etal [Proc. 24th
ICPS, 1083 (1998)]. We have made three-dimensional
(3D) modeling of the strain and the piezoelectric field and
calculated the 3D density of states of strain induced
quantum dots. On the basis of this analysis we have
developed a spin dependent Monte Carlo model, which
describes the carrier dynamics in QD's when the intraband
relaxation is modulated by THz radiation. We show that
THz radiation causes resonance transfer of holes from dark
to radiative states in strain-induced QD's. The transition
includes a spatial transfer of holes from the piezoelectric
potential mimima to the deformation potential minimum. This
phenomenon strongly enhances the QD ground state
luminescence at the expense of the luminescence from higher
states. Our model also reproduces the delayed flash of QD
ground state luminescence, activated by THz radiation even
$\sim1$~s after the carrier generation. Our
simulations suggest a more general possibility to cool the
radiative exciton subsystem in optoelectronic devices.
\end{abstract}
%
\maketitle		  					 	 	  %

\section{Introduction}
Semiconductor quantum dots (QD's) are commonly called
artificial atoms, due to their man-made single-atom-like
spectrum, consisting of well separated spectral
lines.\cite{Kastner1993,Michler2000,Reithmaier2004}
Nevertheless, time-resolved spectroscopy of intraband
relaxation sequences has proved to be much more difficult
with QD's than with free atoms. It is, in distinction to
free atoms, not possible to observe QD intraband
relaxation as sequences of well resolved intermediate
steps, because the intermediate states are very close in
energy and involve many simultaneously excited electrons.

Recently Yusa \etal used visible light and near infrared
(NIR) radiation in combination with terahertz (THz)
radiation to study intraband carrier relaxation in
strain-induced quantum dots (SIQD's).\cite{Yusa1998}
The experiment of Yusa et al, analyzed theoretically in
this work, has remained to our knowledge the only
measurement in which the details of the QD intraband
relaxation processes have been successfully studied by
modulating the dynamics of the carriers, with a radiation
field, which is in approximative resonance with the
intraband transitions. This has prompted us to make a
careful analysis of this experiment in order to emphasize
the potential of this experimental approach and to support
a detailed planning of further experiments of this kind.

In this work we analyze the experiments of Yusa \etal using
a spin-resolved Monte Carlo (MC) carrier dynamics model,
which is based on new full three-dimensional (3D) strain and electronic
structure calculations and furthermore on information
extracted from previous
time-resolved\cite{Grosse1997,Brasken1998} and
continuous wave (CW)\cite{Lipsanen1995} luminescence measurements, including
their computational analysis. We show that deep minima of
the piezoelectric potential (PEP) play a key role in the
observed anomalous THz radiation enhancement of the ground
state luminescence.

The electronic structure was calculated using continuum
elasticity and the multiband \kp~method, which are well
suited for the modeling of relatively large SIQD's (the
ground state wavefunctions have an effective diameter of
approximately $15-20\,$nm).\cite{Wood1996} We have made a
detailed comparison of the theoretical CW
luminescence with experiments and previous axially
symmetric calculations. On the basis of this comparison we
build a simplified electron and hole structure. This was
necessary to make the MC dynamics simulations
computationally feasible.

Our semiempirical carrier dynamics model is based on an
intuitive description of the key physical processes
governing the carrier dynamics of a SIQD. The separate
relaxation processes and their governing rate parameters
are taken from several previous experiments and from
computational analyses of these
experiments.\cite{Grosse1997,Brasken1998} Our MC
model is a generalization of previous master equation models\cite{Grundmann1997,Grosse1997,Brasken1998} by
including the spin of the carriers and by accounting for
the piezoelectric potential minima.

We emphasize that fully \textit{ab initio} calculations of
the studied electron-hole system, including for instance
correlation effects in a nonperturbative way, is at present
not computationally feasible. However, we hope that our
work will prompt for detailed studies of correlation effects
as well as for advanced studies of the electronic structure
using microscopic theories.\cite{Williamson2000}

\section{Electroelasticity and band structure} 				  %
\label{sec:3DModel}                                                               %
%
The understanding of SIQD's has until now relied on a
simplified axisymmetric description of the electronic
structure.\cite{Tulkki1995} In this work we have carried
out extensive three-dimensional modeling of the
electroelastic, electronic and photonic properties of
SIQD's. We emphasize the role of the PEP, not present in
previous two-dimensional models
of the carrier dynamics. Figure~\ref{fig1}(b) shows our
SIQD model geometry, which is based on transmission
electron micrographs of SIQD's.\cite{Georgsson1995}

\subsection{Atomic structure}
The elastic strain and the related piezoelectric potential
were calculated using the finite element method within the
continuum elasticity (CE) approximation. This was motivated
by the large geometry, spanning over $\sim10^8$ atoms and
the possibility to include piezoelectric coupling in the
model. In the coupled electroelastic field formulas,
the vectors of the stress $\mathbf{T}$ and the electric
flux $\mathbf{D}$ are related to the strain $\mathbf{S}$
and the electric field $\mathbf{E}$
 as\cite{Boxberg2005}
\begin{equation}
\label{equ:masterequ}
\begin{array}{cccc}
\left[
\begin{array}{c}
\mathbf{T}\\
\mathbf{D}
\end{array}
\right]
 &
=
&
\left[
\begin{array}{cc}
\mathbf{C}   &  ~\, \mathbf{e} \\
\mathbf{e}^T & -    \boldsymbol{\varepsilon}
\end{array}
\right]
&
\left[
\begin{array}{c}
~\, \mathbf{S} \\
-   \mathbf{E}
\end{array}
\right]~.
\end{array}
\end{equation}
The symmetry of the crystal enters Eq.~\ref{equ:masterequ}
through the elasticity matrix $\mathbf{C}$ and the
piezoelectric matrix $\mathbf{e}$. The dielectric
properties of the materials are described by the dielectric
matrix
$\boldsymbol{\varepsilon}= \varepsilon \mathbbm{1}_{3\times3}$.

The lattice mismatch was introduced into
Eq.~\ref{equ:masterequ} by expanding the different
materials isotropically according to their lattice mismatch
with respect to the GaAs substrate $\Delta a_j = (a_j-a_{GaAs})/a_j$,
where $a_j$ is the lattice constants of either InP or InGaAs
and $a_{GaAs}$ is the lattice constant of GaAs.
Equation~\ref{equ:masterequ} was solved
iteratively,\cite{Ansys} using tetrahedral second order
elements and the CE parameters given in Table~\ref{tab:EC}.

Figure~\ref{fig2} shows the computed strain-induced
electron and hole potentials in the middle of the quantum
well (QW).\cite{Virkkala2000} The total potential is the
sum of the band-edge discontinuity, deformation and
piezoelectric potentials. The elastic strain gives rise to
the main deformation potential (DP) minima in the center.
The PEP gives, on the contrary, rise to large side minima
and side barriers of holes, located near the edges of the
InP island. The effect of the PEP on the center minima is,
however, small.

\subsection{Electronic structure}
\label{sec:electron}
The electronic structure of the QD's was calculated using
the eight-band ${\bf k} \cdot {\bf p}$
method,\cite{Pryor1998,Stier,Singh1998} accounting for band
edge discontinuity, strain-induced deformation potentials
and the piezoelectric potential. The eight-band
${\bf k} \cdot {\bf p}$ Hamiltonian can be divided in a
strain-free part and a strain part. The strain-free part is
parametrized by the band gap $E_g$, the spin-orbit
splitting energy $\Delta$, the optical matrix element $E_p$,
the valence band offset $VBO$, the modified Luttinger
parameters $\gamma_1$, $\gamma_2$ and $\gamma_3$, the
conduction band effective mass $m_e$ and Kane's
parameter $B$. The strain part, on the other hand, is
parametrized by the deformation potentials $a_c$, $a_v$,
$b_v$ and $d_v$. Moreover, it contains the six components
of the symmetrized strain tensor $\varepsilon_{xx}$,
$\varepsilon_{yy}$, $\varepsilon_{zz}$, $\varepsilon_{xy}$,
$\varepsilon_{yz}$, $\varepsilon_{xz}$, and the
piezoelectric potential $\phi_{piez}$.\cite{Georgsson1995}

The eight-band ${\bf k} \cdot {\bf p}$ Hamiltonian was
discretized using the finite difference technique and
diagonalized using the implicitly restarted Lanczos
method\cite{Arpack}. For more details on the electronic
structure calculations see Ref. \onlinecite{Boxberg2005}.
The material parameters used in the eight-band
${\bf k} \cdot {\bf p}$ calculations are listed in
Table~\ref{tab:Mat}. The value of the conduction-valence
band coupling parameter $E_p$, given in
Table~\ref{tab:Mat}, has been reduced in order to avoid
spurious states (see Ref. \onlinecite{Boxberg2005}, and
references therein). However, the effect of the reduction
of $E_p$ on the effective masses was negligible (this was
verified by testing the model for the pertinent bulk
semiconductors).

The depth of the PEP minima (+65~meV, see Fig.~\ref{fig2})
of holes exceeds that of the DP minimum (+24 meV) by tens
of meV. The PEP minima have consequently many confined hole
states and it is, therefore, energetically more favorable
for the holes to populate the deepest PEP states before
occupying the DP states. The electrons are, however, mainly
confined in the DP minimum (-108~meV), which over-rules the
shallow PEP minima of electrons (-15~meV). Spatially
separated electron (located on the DP ground state $e_1$)
and hole (on the PEP level $p_1$) pairs have several
decades smaller probability of recombining radiatively than
an exciton confined to the DP levels $e_1$ and $h_1$.

Figure~\ref{fig3} shows the probability densities of the
lowest optically active electron and hole states, confined
in the DP minima in the center of the SIQD.
Figure~\ref{fig4} shows the density of states (DOS) of all
(bright and dark) confined states, including also the hole
states confined in the PEP minima (dark). The eigenenergies
of Fig.~\ref{fig4} are given with respect to the strain-free
band edge energy of the QW material.

\subsection{Photoluminescence}
\label{photon}
The PL intensity was calculated by summing over the
transition rates from all electron states to all hole
states. The differential photoluminescence rate (rate of
spontaneous emission of photons) of light with the
polarization $\boldsymbol{\alpha}$ is given by
\begin{equation}
\label{equ:PL}
\begin{split}
I_{\boldsymbol{\alpha}}(E) \propto
 \sum\limits_{i,f} \frac{g(E)}{E}
\left|
\overlap{\Psi_{ei}}{\boldsymbol{\alpha}^*\cdot \p}{\Psi_{hf}}
\right|^2
f_e(E_i) \, f_h(E_f) \, P \left\{ [E-\left(E_i - E_f  \right)] ,\Gamma \right\} ~,
\end{split}
\end{equation}
where $g(E)$ is the density of photon states
[$g(E)\propto E^2$ for a 3D system, i.e., in the absence of
a cavity], $E_i$ and $E_f$ are the eigenenergies of the
initial electron states $\ket{\Psi_{ei}}$ and the final
hole states $\ket{\Psi_{hf}}$. The inhomogeneous linewidth
broadening of the transition energies
$E_{\hbar\omega}=E_i - E_f$, due to the size distribution
of the SIQD, was approximated using the Gaussian broadening
function $P ( E ,\Gamma )$,\cite{Grundmann1997II,Qasaimeh2003} where $\Gamma$ is the inhomogeneous linewidth. The Lorentz shaped
homogenous linewidth broadening (due to the finite life
times of the QD eigenstates) was neglected, because it has
a vanishing effect on the PL, which is also
inhomogeneously broadened. In the calculations of
Fig.~\ref{fig5} we have assumed complete state filling,
i.e., the Fermi functions of electrons and holes were
$f_e(E_i)=1$ and $f_h(E_f)=1$, respectively. On the basis
of the experimentally observed size distribution of InP
stressor islands\cite{Bryant} we have estimated
$\Gamma=6\,$meV.

Figure~\ref{fig5} shows the simulated PL spectrum, averaged
over the polarization, for an ensemble of SIQD's together
with an experimental luminescence spectrum.
Figure~\ref{fig5} shows also the effect of the exact InP
island geometry and of the PEP on the PL spectrum. All
results are based on fully three-dimensional simulations.
However, in the axisymmetric model we used an InP island
having the shape of a truncated cone with the same height,
the same InP volume and the same bottom area as the angular
island, shown in Fig.~\ref{fig1}(b). We note that the PEP
affects mainly the PL peaks corresponding to transitions
between excited states (QD3, QD4, QD5), whereas the angular
island geometry (the deviation from the axisymmetry)
increases the peak-peak energy separation and sharpens the
peak profiles.

The calculation of radiative transition energies and line
strengths shows good agreement with the experimental
intensity distribution and energy spacing of the
luminescence peaks.\cite{Lipsanen1995} Each peak consists of
many different transitions which after the convolution of
inhomogeneous broadening form a smooth and continuous
spectrum, well reproducing the experimental PL. A
characterization of the PL peaks, according to the symmetry
of the involved eigenstates, is difficult due to the weak
confinement and small energy spacing ($1-4~$meV) of hole
states. Only the two lowest peaks (QD1 and QD2) are due to
recombination between clearly confined
\textit{electron and hole} states. The higher PL peaks are
due to recombination between a few confined electron states
and several confined or \textit{quasiconfined} hole
states (i.e. the hole states are \textit{shared} between
the DP and PEP minima, with their energies merging with the
\textit{energy continuum} of the QW).

\section{Dynamical model}							  %
The presence of the PEP minima makes the relaxation
pathways decisively different from conventional QD carrier
dynamics, especially for holes, for which the piezoelectric
side barriers (shown in Fig.~\ref{fig2}) block the
relaxation to the DP minimum along the $y$ axis. The
relaxation of holes along the $x$ axis to the DP minimum is
hindered as well, but by the PEP minima which capture the
holes before they can enter the DP minima. The inefficient
hole relaxation was also observed in the experiments as
weak QD luminescence during Ti-Sap carrier generation
directly to the QW. This indicated a very inefficient
carrier relaxation from the QW to the radiative states of
the SIQD's unless the carriers of the QW were \textit{hot},
as in the case of Ar-ion laser pumping.

On the basis of these observations we have developed the
carrier dynamics model, schematically shown in
Fig.~\ref{fig6}. This model is a generalization, based on a
true three-dimensional description of the QD system, of the
models presented in Refs. \cite{Grundmann1997} and
\cite{Grosse1997}, which are as such not able to describe
the experiments analyzed here.

\subsection{Underlying assumptions}
Our carrier dynamics model is based on the following
assumptions, which are motivated by both experimental
observations and theoretical studies:

(i) \textit{The carrier population is charge neutral.}
This implies that the total number of holes (including the
valence states of the DP minimum, the two PEP minima and
the QW) is equal to the total number of electrons
(including the conduction states of the DP minimum and
the QW). This is motivated for undoped samples were the
optical carrier generation and recombination conserves the
net charge, except for small charge fluctuations between
different QD's, which mainly broaden the PL peaks. The
assumption of charge neutrality is also supported by the
invariance of the experimentally measured line energies in
SIQD luminescence spectra, as a function of state filling;
i.e., the luminescence energies are independent on the
filling of eigenstates\cite{Lipsanen1995,Lipsanen1996}. If
there was a change in the net charge, there would also be a
noticeable change in the line energies.

(ii) \textit{The total number of electron and hole states,
included in the model, were equal.} The number of electron
and hole states was limited to $N \sim 100$, in order to
stabilize the MC simulations and to account for
the possible saturation of carrier generation to the QW
reservoirs. This was done by setting the total number of
electron states equal to the total number of hole states of
the entire model. This has little effect on the QD
luminescence character as we are typically far from
saturation, but this is in accordance with assumption (i).

(iii) \textit{We excluded the possibility of nonradiative
recombination channels.} Nonradiative recombinations are
typically a result of crystal imperfections or interface
states and they are indirectly observed in terms of very
short luminescence lifetimes. The self-organized samples of
SIQD's are of very high quality, showing very long
luminescence life times of $\sim 1\,$ns.\cite{Grosse1997}
The non-radiative recombination channels are thus rare.

(iv) \textit{Nearly degenerate electron and hole states
were combined to a few degenerate levels.} The DOS had to
be simplified in order to make the MC calculations
numerically feasible. The original DOS contained too many
states to be included in the MC model. However,
the simplified DOS resembles well the true multiband
three-dimensional DOS.

(v) \textit{We accounted for the spin of the carriers.} The
preservation of the spin quantum number influences the
total speed of recombination. The conservation of the spin
should be accounted for in any carrier dynamics model,
however, it has to our knowledge not been accounted for in
previous QD relaxation models. Note, that the spin is in
general not a good quantum number for states of the
eight-band effective-mass Hamiltonian, since the valence
bands are strongly coupled by the spin-orbit interaction.
In the present case, the spin is, however, an approximately
good quantum number for both electron and hole states. This
is due to the vertical confinement and biaxial strain of
the QW, which cause the most strongly confined hole
states to be nearly pure heavy-hole states, which have a
well defined spin.

(vi) \textit{The intraband relaxation time $\tau_{0}$ was
assumed ten times smaller than the radiative life times}.
An equal intraband relaxation time of both electrons and
holes is motivated for high carrier densities in which the
carrier relaxation is dominated by Coulomb scattering and
the Auger process. \cite{Grosse1997,Brasken1998} From
experiments\cite{Grosse1997,Brasken1998} we know that
at relatively high state filling the intraband relaxation
is one order of magnitude faster than the radiative
recombination. Furthermore, we assumed that relaxation
takes place between consecutive states only.

(vii) \textit{The tunneling rate between the PEP and DP
minima was assumed zero.} This tunneling rate is small in
comparison with the prevailing relaxation and recombination
rates, since the spatial separation of the DP and PEP
minima is about $20\sim30\,$nm.

(viii) \textit{It was assumed that an intense THz radiation
can couple the hole states $p_1$ and $h_1$.} If the
intensity of the THz radiation is large and if the THz
photon energy matches the energy separation of $h_1$ and
$p_1$ it gives rise to a resonant charge transfer where the
absorption (hole: $p_1 \rightarrow h_1$) and emission
(hole: $h_1 \rightarrow p_1$) of THz photons compete. The
radiation field is assumed to be in resonance with the
$h_1$ DP state and \textit{any} of the deep PEP states,
which are in the model described by a single degenerate
level $p_1$. We neglect spontaneous emission, which is
presumably small at high THz intensities. The strength
of the hole transfer is, in the experiment, not limited to
a very narrow THz frequency range since in
the real SIQD sample the single resonance frequency of the
model is replaced by a \textit{quasi-continuum of
transition frequencies}. The inhomogeneous size
distribution of the QD's is likely to further broaden the
otherwise narrow peaks of allowed THz radiation frequencies
of effective coupling.

The coupling strength between the PEP states $p_1$ and the
DP states $h_1$ depends on the intensity of the THz field,
the carrier populations of the involved hole levels and a
phenomenological coupling constant. This coupling strength
is the only free scaling constant in our model. The
available experimental and numerical tools are currently
far from enabling \textit{quantitative} measurements or
\textit{ab initio} type calculations of this coupling
constant.
As a summarizing note on our relaxation model we recall
that all model parameters, except the phenomenological
THz radiation coupling, are based on direct measurements
($\tau_{ri}$, $\tau_0$), fitting to experimental data
($f_i$), or electron structure calculations
($g_i$).\cite{Lipsanen1995,Grosse1997,Brasken1998}

\subsection{Master equations}
The time evolution of the electron ($n_i^e$) and hole
($n_i^h$) populations of the intermediate DP states are
described by the following master equations for
$2 \leq i \leq 4$,
\begin{align}
\label{equ:dyn1}
\frac{dn_i^e}{dt} & 	= \frac{n_{i+1}^e\left(g_i-n_i^e\right)}{\tau_0}
			- \frac{n_{i}^e\left(g_{i-1}-n_{i-1}^e\right)}{\tau_0}
			- \frac{n_i^e n_i^h }{g_i \tau_{ri}} f_i\notag\\
&			- \frac{n_i^e [n_{(i+1)A}^p+n_{(i+1)B}^p] }{2 g_i \tau_{ri}}
			  (1-f_i), \\
\label{equ:dyn2}
\frac{dn_i^h}{dt} & 	= \frac{n_{i+1}^h\left(g_i-n_i^h\right)}{\tau_0}
			- \frac{n_i^h\left(g_{i-1}-n_{i-1}^h\right)}{\tau_0}
			- \frac{n_i^e n_i^h }{g_i \tau_{ri}} f_i
\end{align}
and for the PEP states with $2 \leq i \leq 5$
\begin{align}
\label{equ:dyn3}
\frac{dn_{i}^p}{dt} &  	= \frac{n_{i+1}^p\left(g_i^p-n_i^p\right)}{\tau_0}
			- \frac{n_i^p\left(g_{i-1}^p-n_{i-1}^p\right)}{\tau_0}\notag\\
&			- \frac{n_{i-1}^e n_i^p }{2 g_i \tau_{ri}} (1-f_{i-1}).
\end{align}
The time evolution of the ground state ($i=1$) populations are
given by
\begin{align}
\label{equ:dyn4}
\frac{dn_1^e}{dt} & 	= \frac{n_2^e\left(g_1-n_1^e\right)}{\tau_0}
			- \frac{n_1^e n_1^h }{g_1 \tau_{r1}} f_1
			- \frac{n_1^e \left[n_{2A}^p+n_{2B}^p\right]}{2g_1 \tau_{r1}}(1-f_1),\\
\label{equ:dyn5}
\frac{dn_1^h}{dt} & 	=  \frac{n_2^h\left(g_1-n_1^h\right)}{\tau_0}
			- \frac{n_1^e n_1^h }{g_{r1} \tau_{r1}}f_1 \notag \\
&			+  \left[g_1 \left(n_{1A}^p+n_{1B}^p\right) -2n_1^hg_1^p\right]G_{THz}, \\
\label{equ:dyn6}
\frac{dn_{1\gamma}^p}{dt} & 	=  \frac{n_{2\gamma}^p\left(g_1^p-n_{1\gamma}^p\right)}{\tau_0}
			+  \left(n_1^h g_1^p - n_{1\gamma}^p \, g_1^p\right) G_{THz},
\end{align}
and for the reservoirs ($i=R$) we have
\begin{align}
\label{equ:dyn7}
\frac{dn_R^e}{dt} & 	=
			- \frac{n_R^e\left(g_4-n_4^e \right)}{\tau_{eQD}}
			- \frac{n_R^e n_R^h }{g_R \tau_{R}}
			+ G, \\
\label{equ:dyn8}
\frac{dn_R^h}{dt} & 	=
			- \frac{n_R^h\left(g_4-n_4^h\right)}{\tau_{hQD}}
			- \frac{n_R^h\left(2g_5^p-n_{5A}^p-n_{5B}^p\right)}{\tau_{pQD}} \notag \\
&			- \frac{n_R^e n_R^h }{g_R \tau_{R}}
			+ G,
\end{align}
where the two PEP minima are identified by subindices
$\gamma=iA$ and $\gamma=iB$. In Eqs.~\ref{equ:dyn1}-\ref{equ:dyn8}, is the THz radiation
intensity and
$G=\left(g_R^e-n_R^e\right)\left(g_R^h-n_R^h\right)G_{exc}$
is the carrier generation rate by the pump laser $G_{exc}$.
We also accounted for the spin of the carriers, which
effectively raised the population of excited states.
A detailed analysis of Eqs.~\ref{equ:dyn1}-\ref{equ:dyn8}
follows below.

\subsection{Partial transition rates}
The MC carrier dynamics model is depicted in
Fig.~\ref{fig6} and was given in mathematical form in
Eqs.~\ref{equ:dyn1}-\ref{equ:dyn8}. It describes the
time evolution of the electron $n_i^e$ and hole $n_i^h$
populations of the DP states as well as the hole
populations $n_{iA}^p$ and $n_{iB}^p$ of the PEP states
(left and right minima).

\subsubsection{The population of intermediate states}
The time evolution of the electron and hole populations
$n_i^e$ and $n_i^h$ of the intermediate states
(i.e., $2\leq i \leq 4$) of the DP minima are described by
Eqs.~\ref{equ:dyn1} and ~\ref{equ:dyn2}.
The population of state $i$ is increased by the intraband
relaxation (colored/gray arrows in Fig.~\ref{fig6}) from
state $i+1$ to state $i$
[first term on the right hand side (RHS) of
Eqs.~\ref{equ:dyn1} and \ref{equ:dyn2}] and reduced by the
carrier relaxation to state $i-1$
[second term on the RHS of Eqs.~\ref{equ:dyn1} and
\ref{equ:dyn2}].
The rate of the intraband relaxation is described
by the intraband relaxation lifetime $\tau_0$, which is
typically one tenth of the radiative lifetime
$\tau_r$.\cite{Brasken1998}

The main recombination processes involve electron and
holes (the center black dashed arrow in Fig.~\ref{fig6}),
both confined in the DP minima, and are described by the
third term on the RHS of Eqs.~\ref{equ:dyn1} and
\ref{equ:dyn2}. The recombination of electrons in the DP
minimum and holes in the PEP minima is described by the
fourth term on the RHS of Eq.~\ref{equ:dyn1} (left and
right, black dashed arrows in Fig.~\ref{fig6}). This rate
is weaker than the recombination rate between the DP states,
due to the small spatial overlap of the pertinent wave
functions. The different transition probabilities of the
DP$\rightarrow$DP and DP$\rightarrow$PEP recombination
could be included by using different time constants.
However, in our model it has been more convenient to
describe this effect by the ratios $f_i/(1-f_i)$.
The numerical values of $f_i$ were
determined by fitting the state filling (PL peak
intensities), as a function of the generation rate, to
experimental PL data.\cite{Lipsanen1995,Lipsanen1996}

The time evolution of the hole populations $n_i^p$ of the
intermediate (i.e. $2 \leq i \leq 5$) states of the PEP
minima are described by Eq.~\ref{equ:dyn3}. The first and
second terms on the RHS correspond to
intraband relaxation (colored/gray arrows in
Fig.~\ref{fig6}) from state $i+1$ to state $i$ and from
state $i$ to $i-1$, respectively. The last term
corresponds to radiative recombination with electrons
confined in the DP minimum (left and right, black dashed
arrows in Fig.~\ref{fig6}).

\subsubsection{The population of the DP and PEP ground states}
The time evolution of the electron ground state population
$n_1^e$ is described by Eq.~\ref{equ:dyn4}, which is
similar to Eq.~\ref{equ:dyn1}, except that it does not
contain the second term of the RHS of Eq.~\ref{equ:dyn1} as
the holes cannot relaxate any further from the ground
state.
Equation \ref{equ:dyn5} describes the time evolution of
the hole population of the DP ground state. The first and
second terms on the RHS correspond to intraband relaxation
from state $i+1$ to state $i$ (colored/gray arrows in
Fig.~\ref{fig6}) and to radiative recombination (center
black dashed arrow in Fig.~\ref{fig6}) between electron and
holes of the DP minima. The third term correspond
to the THz radiation-induced coupling of the states $h1$,
$p1A$ (in the left PEP minimum) and $p1B$ (in the right
PEP minimum). It accounts for the absorption of a
THz photon, exciting a hole from state $h1$ to states $p1A$
or $p1B$, and the stimulated emission of a THz photon,
taking a hole from state $p1A$ to $h1$ or from state $p1B$
to $h1$. These processes are indicated by double-headed
arrows in Fig.~\ref{fig6}.

The time evolution of population $n_{1\gamma}^p$ of the
deepest PEP states is described by Eq.~\ref{equ:dyn6}, where
$\gamma=A$ or $\gamma=B$ correspond to left and right PEP
minimum, respectively. The first term on the RHS
corresponds to the intraband relaxation from state $p2$ to
state $p1$ (colored/gray arrows in Fig.~\ref{fig6}). The
second term accounts for both the emission or absorbtion of
a THz photon (simultaneously moving a hole from state
$p1\gamma$ to state $h1$).

\subsubsection{The population of the reservoirs}
The physical QW, embedding the SIQD's is modelled as a
reservoir of electrons and holes, with the state
degeneracies $g_R^e$ and $g_R^h$, both being much larger
than the degeneracies of the QD levels. The electron and
hole reservoirs are described by Eqs.~\ref{equ:dyn7} and
\ref{equ:dyn8}, respectively. The first term on the RHS of
Eq.~\ref{equ:dyn7} [Eq.~\ref{equ:dyn8}] describes the
capturing of an electron (hole) from the reservoirs to the
DP minimum with the capture lifetime $\tau_{eQD}$
($\tau_{hQD}$) describing the rate of this process. The
second term on the RHS of Eq.~\ref{equ:dyn7} and the third
term on the RHS of Eq.~\ref{equ:dyn8} account for the
radiative recombination of electrons and holes in the
reservoirs, with the corresponding lifetime $\tau_R$.
The second term of Eq.~\ref{equ:dyn8} is due to holes
getting captured into the two ($A$ and $B$) PEP minima.
The carrier generation is described by the last term of
both Eq.~\ref{equ:dyn7} and \ref{equ:dyn8}.

\subsection{General remarks on the model parametrization}
We note that the radiative lifetimes of SIQD's have been
measured and modeled successfully
previously.\cite{Grosse1997,Brasken1998} In our carrier
dynamics model we used radiative lifetimes $\tau_{ri}$
($\tau_{ri} > \tau_{rj} \approx 10 \, \tau_0$, for all
$i>j$), which were calculated from the eight-band electron
structure model in the electric dipole approximation and
were in good agreement with experiments. The radiative
recombination rates of the $p1$ states were found
negligible. We found that the radiative recombination
rates of also the lower excited PEP states ($pi$, with
$i\lesssim 4$) were noticeable longer than those of the
corresponding DP states. We used, therefore, recombination
lifetimes $\tau_{ri}^\prime = \tau_{ri}/(1-f_i)$ with
$0<f_i<1$ for the excited states of the PEP minima. The
exact values of $f_i$ were determined by fitting the state
filling (PL peak intensities) to experimental
data.\cite{Lipsanen1996} The distinction between DP and PEP
states becomes, however, difficult for highly excited
states, since these states are not confined in only one
minimum, but the probability densities are shared between
all three minima.

The intraband relaxation time constant $\tau_0$ affects,
together with the amplitudes of the individual
recombination processes, the relative CW intensities and
the rates of fading off of the PL peaks. The effect of the
THz radiation on the QD PL is not very sensitive to
$\tau_0$ at low generation intensities. However, the
smaller $\tau_0$ is, the smaller generation intensity is
needed to saturate the hole population of $h1$ (even
without any THz radiation). The value of $\tau_0$ does
thereby affect the threshold value of the generation
intensity at which the THz radiation-induced PL effect
disappears.

The validity of the used radiative and intraband
relaxation time constants, describing the carrier dynamics
\textit{in the absence} of any THz radiation, was verified
by comparing the numerical time dependent PL of our model
with experiments. On the basis of this comparison we
conclude that our model is very well in line with
Refs.~\cite{Lipsanen1996} and \cite{Grosse1997} (see also
Fig.~\ref{fig5}).

Equations \ref{equ:dyn1}-\ref{equ:dyn8} were solved by
time-dependent MC simulations\cite{MC} of a large ensemble
of independent QD's. We obtained the probabilities of each
many-particle partition as a function of time by averaging
over the ensemble populations. By simulating QD ensembles
large enough (a few thousands of QD's) we were able to
minimize the time-dependent oscillations of simulated
\textit{quasistationary} partitions during steady state
laser modulation.

\section{Discussion and conclusions}
\subsection{Luminescence during continuous pumping}				  %
The free carriers can relax from the GaAs barrier to the
DP and PEP minima,
either directly or through the QW state continuum. The
direct relaxation is assisted by the funnel shaped
deformation and piezoelectric potential minima in the
barriers.\cite{Tulkki1995} As the holes are predominantly
confined in the deep PEP minima, the radiative
recombination starts only when these minima are already
filled with holes. A saturated QD is consequently strongly
polarized during continuous generation and recombination.
The charge separation, with holes in the PEP minima and
electrons in the DP minimum, can persist even seconds after
turning off the carrier generation.

The picture described above does not account for any
excitonic carrier-carrier interactions. Previous studies
have, however, shown that the carrier-carrier correlation
effects are small in SIQD's.\cite{Brasken2001} The very
good qualitative agreement between the experimental CW PL
and the PL simulated here does also support the used
single-particle approximation. The exciton effects of
SIQD's are, though, fairly vaguely understood and do
call for further studies.

\subsection{Influence of THz radiation on the steady-state luminescence}	  %
According to our simulations, the enhancement of the QD1
luminescence is due to a THz radiation induced continuous
drift of holes between the PEP $p_1$ levels and the DP
$h_1$ level (Fig.~\ref{fig8}). Under steady-state condition
the \textit{total} integrated PL intensity is conserved
\textit{in the first approximation}. The QD2, QD3, QD4, and
QW peaks are accordingly reduced when the QD1 peak is
increased by the THz radiation, both in the experiment and
theory. Our calculations predict furthermore that the
THz radiation enhancement of the QD1 peak disappears at
high Ar$^+$ pumping intensities as a result of the
saturation of the hole population at $h_1$.

The results are in good qualitative agreement with the
experiments, although, a quantitative agreement cannot be
achieved with the current model. Our model does not show as
dramatic quenching of the QW PL by the THz radiation as was
seen in the experiment, although, Figs.~\ref{fig7} and
\ref{fig8} do show a clear decrease of the QW PL. We argue
that the large decrease of the experimental QW PL is
related to the heating and ionization of QW
excitons.\cite{Cerne1996}

The discrepancies between the intensity distributions of
the PL peaks (in Fig.~\ref{fig7}) of our model and of the
experiments are mainly due to the simplified electronic
structure used in our MC model and the fact that
we have used the same relaxation life time for all hole and
electron states. A better fit between the simulations
and experiments could have been obtained by using,
\textit{e.g.}, several different relaxation rates
between the different hole levels. This would, however,
have made the model less transparent. The exact PL intensity distribution between the different
peaks is also very much dependent on the carrier generation
rate. This is true for both the simulations and the
experiments (see, \textit{e.g.}, Ref.~\cite{Grundmann1997}).
Figure 4 was plotted for a carrier generation rate where
the simulated QD2 peak already exceeds the QD1 peak. This
generation rate was chosen, in order to obtain as clear THz
effect as possible.

\subsection{THz radiation-induced transients}             	                  %
Figures \ref{fig8}(a) and \ref{fig8}(b) show experimental and
simulated time-resolved PL. The black and red curves
correspond to integrated PL of the QD1 and QW peaks,
respectively. The left panels show results obtained for
overlapping time windows of Ti-Sap pumping and
THz radiation excitation. The QW peak is decreased and the
QD1 PL is increased during the THz radiation with both PL
peaks returning to their initial value after turning off
the THz radiation. Note the saturation of both the
experimental and theoretical PL during THz radiation.

\subsection{THz radiation-induced delayed PL}					  %
The right panels of Fig.~\ref{fig8} show PL with a delayed
THz radiation pulse after turning off the pump laser. The
rise of the QD1 peak by the delayed THz radiation is due to
the excitation and drift of trapped holes from the PEP
minima to the DP minimum, where they recombine with
electrons localized in the DP minimum. The simulation time
was much smaller than that of the experimental time scale,
due to computational limitations. However, the time scale
of the simulations were chosen long enough to observe
saturation of the PL both during the pumping and between
the pumping and the delayed THz radiation. There is also in
this case a very good qualitative agreement between
experiments and theory.

\subsection{Conclusions}							  %
We have shown that the anomalous cooling of radiative QD
excitons\cite{Yusa1998} is due to a THz radiation-induced
drift of holes from piezo-electric potential (PEP) minima
to the lowest radiative states of the QD. The subsystem of
radiative excitons is cooled, although, the average energy
of the total carrier system increases (heated).

By cooling of the \textit{radiative exciton subsystem} we
mean that the average hole population of the ground
state $h1$ increases whereas the average population of all
excited states in the deformation potential (DP) minimum
decreases. The large shift
of the particle population (within the radiative states)
from the excited states towards the ground state is
analogous to lowering of the temperature in a system of
thermal equilibrium. However, the carriers of our SIQD's
are not in thermal equilibrium nor in thermal
quasi-equilibrium during continuous (or transient) carrier
generation and THz modulation. Therefore, the reduction of
the average energy of the radiative excitons cannot be
called lowering of the carrier temperature.

Our computational model reproduces currently the following
experimental observations on SIQD's.
(i) The simulated CW PL reproduces experimental
data\cite{Lipsanen1996} as a function of the carrier
generation intensity.
(ii) The THz radiation increases the ground state
luminescence and decreases the excited state luminescence
at low generation intensities during CW THz excitation.
(iii) At the onset of the THz radiation there is a strong
and sudden increase of the QD1 luminescence. The
enhancement by the THz radiation is then, however,
exponentially reduced and saturates at a lower level.
(iv) The THz radiation gives rise to ground state
luminescence long after turning off the carrier generation.
This delayed PL flash is seen only in the ground state
luminescence and it fades out exponentially.

The ultimate goal of our work is to motivate new
experiments in which the THz photon energy would be varied
over a wide energy range, covering several $pi$ and
$hi$ levels. We argue that the analysis of the
original experiments by Yusa \textit{et al.} is necessary
for a careful planning of further experiments.
Finally, we note that our simulations suggest that the
resonance transfer by THz radiation could be replaced by a
bias voltage-driven resonance tunneling, leading to an
enhanced QD ground state luminescence.

\section*{Acknowledgements}
We gratefully acknowledge the fruitful communications with
Harri Lipsanen and the computational effort by Roman
Terechonkov on the electronic structure calculations.

\begin{thebibliography}{10}

\bibitem{Ansys}
All elasticity calculations were performed using the commercial software Ansys.

\bibitem{LB}
Numerical data and functional relationships in science and technology.
\newblock Berlin, Germany: Springer (1991) (Landolt-B{\"o}rnstein. New Series,
  III/17a).

\bibitem{Brasken1998}
M.~Brask\'en, M.~Lindberg, M.~Sopanen, H.~Lipsanen, and J.~Tulkki.
\newblock Temperature dependence of carrier relaxation in strain-induced
  quantum dots.
\newblock {\em Phys. Rev B}, 58:R15993--R15996, 1998.

\bibitem{Brasken2001}
M.~Brask{\'e}n, M.~Lindberg, D.~Sundholm, and J.~Olsen.
\newblock Spatial carrier-carrier correlations in strain-induced quantum dots.
\newblock {\em Phys. Rev B}, 64:035312, 2001.

\bibitem{Cerne1996}
J.~Cerne, J.~Kono, M.~S. Sherwin, M.~Sundaram, A.~C. Gossard, and G.~E.~W.
  Bauer.
\newblock Terahertz dynamics of excitons in {GaAs/AlGaAs} quantum wells.
\newblock {\em Phys. Rev Lett.}, 77:1131--1134, 1996.

\bibitem{Georgsson1995}
K.~Georgsson, N.~Carlsson, L.~Samuelson, W.~Seifert, and L.~R. Wallenberg.
\newblock Transmission electron microscopy investigation of the morphology of
  inp stranski-krastanow islands grown by metalorganic chemical vapor
  deposition.
\newblock {\em Appl. Phys. Lett.}, 67:2981--2982, 1995.
\newblock The {InP} stressors of this work were similar to those of the
  presently studied sample, however, these samples did not contain any QW.

\bibitem{Grosse1997}
S.~Grosse, J.~H.~H. Sandmann, G.~von Plessen, J.~Feldmann, H.~Lipsanen,
  M.~Sopanen, J.~Tulkki, and J.~Ahopelto.
\newblock Carrier relaxation dynamics in quantum dots: {S}cattering mechanisms
  and state-filling effects.
\newblock {\em Phys. Rev B}, 55:4473--4476, 1997.

\bibitem{Grundmann1997II}
M.~Grundmann and D.~Bimberg.
\newblock Gain and threshold of quantum dot lasers: {T}heory and comparison to
  experiment.
\newblock {\em Jpn. J. Appl. Phys.}, 36:4181--4187, 1997.

\bibitem{Grundmann1997}
M.~Grundmann and D.~Bimberg.
\newblock Theory of random population for quantum dots.
\newblock {\em Phys. Rev B}, 55:9740--9745, 1997.

\bibitem{Singh1998}
H.~Jiang and J.~Singh.
\newblock {\em IEEE J. Quantum Electron.}, 34:1188, 1998.

\bibitem{Kastner1993}
M.~A. Kastner.
\newblock Artificial atoms.
\newblock {\em Physics Today}, 46:24, 1993.

\bibitem{Lipsanen1995}
H.~Lipsanen, M.~Sopanen, and J.~Ahopelto.
\newblock Luminescence from excited states in strain-induced
  {I}n$_x${G}a$_{1-x}${A}s quantum dots.
\newblock {\em Phys. Rev B}, 51:13868--13871, 1995.

\bibitem{Lipsanen1996}
H.~Lipsanen, M.~Sopanen, and J.~Ahopelto.
\newblock Fabrication and photoluminescence of quantum dots induced by strain
  of self-organized stressors.
\newblock {\em Solid-State Electron.}, 40:601--604, 1996.

\bibitem{Bryant}
H.~Lipsanen, M.~Sopanen, and J.~Tulkki.
\newblock {\em Optics of quantum dots and wires}, pages 97--131.
\newblock Artech House, 2005.

\bibitem{Michler2000}
P.~Michler, A.~Kiraz, C.~Becher, W.~V. Schoenfeld, P.~M. Petroff, Lidong Zhang,
  E.~Hu, and A.~Imamoglu.
\newblock {A Quantum Dot Single-Photon Turnstile Device}.
\newblock {\em Science}, 290:2282--2285, 2000.

\bibitem{Pryor1998}
C.~Pryor.
\newblock Eight-band calculations of strained {I}n{A}s/{G}a{A}s quantum dots
  compared with one-, four, and six-band approximations.
\newblock {\em Phys. Rev B}, 57:7190--7195, 1998.

\bibitem{Qasaimeh2003}
O.~Qasaimeh.
\newblock Effect of inhomogeneous line broadening on gain and differential gain
  of quantum dot lasers.
\newblock {\em {IEEE} Trans. Electron Devices}, 50:1575--1581, 2003.

\bibitem{Reithmaier2004}
J.~P. Reithmaier, G.~Sek, A.~L{\"o}ffler, C.~Hofmann, S.~Kuhn, S.~Reitzenstein,
  L.~V. Keldysh, V.~D. Kulakovskii, T.~L. Reinecke, and A.~Forchel.
\newblock Strong coupling in a single quantum dot-semiconductor microcavity
  system.
\newblock {\em Nature}, 432:197--200, 2004.

\bibitem{Singh}
J.~Singh.
\newblock {\em Physics of Semiconductors and Their Heterostructures}.
\newblock Mc Graw-Hill, 1993.

\bibitem{Arpack}
Society for Industrial \& Applied Mathematics.
\newblock {\em ARPACK Users' Guide: Solution of Large-Scale Eigenvalue Problems
  with Implicitly Restarted Arnoldi Methods}, 1998.

\bibitem{Stier}
O.~Stier.
\newblock {\em Electronic and Optical Properties of Quantum Dots and Wires}.
\newblock Wissenschaft und Technik Verlag, 2001.

\bibitem{Tulkki1995}
J.~Tulkki and A.~Hein{\"a}m{\"a}ki.
\newblock Confinement effect in a quantum well dot induced by an {InP}
  stressor.
\newblock {\em Phys. Rev B}, 52:8239--8243, 1995.

\bibitem{Boxberg2005}
S.~v~Alfthan, F.~Boxberg, K.~Kaski, A.~Kuronen, R.~Tereshonkov, J.~Tulkki, and
  H.~Sakaki.
\newblock Electronic, optical, and structural properties of quantum wire
  superlattices on vicinal (111) {GaAs} substrates.
\newblock {\em Phys. Rev B}, 72:045329, 2005.

\bibitem{Virkkala2000}
R.~Virkkala, K.~Maijala, and J.~Tulkki.
\newblock Piezoelectric potentials and carrier lifetimes in strain-induced
  quantum well dots.
\newblock {\em Phys. Rev B}, 62:6932--6935, 2000.
\newblock The results reported here differ from the earlier ones due to a
  decisively improved electro-elastic model and more rigorous accouting of the
  crystal and island symmetry.

\bibitem{Vurgaftman2001}
I.~Vurgaftman, J.~R. Meyer, and L.~R. Ram-Mohan.
\newblock Band parameters for {III-V} compound semiconductors and their alloys.
\newblock {\em J. Appl. Phys.}, 89:5815--5875, 2001.

\bibitem{MC}
C.~Wasshuber.
\newblock {\em Computational Single-Electronics}.
\newblock Computational Microelectronics. Springer, 2001.
\newblock See \textit{eg.} chapter 3 of.

\bibitem{Williamson2000}
A.~J. Williamson, L.~W. Wang, and A.~Zunger.
\newblock Theoretical interpretation of the experimental electronic structure
  of lens-shaped self-assembled {InAs/GaAs} quantum dots.
\newblock {\em Phys. Rev B}, 62:12963--12977, 2000.

\bibitem{Wood1996}
D.~M. Wood and A.~Zunger.
\newblock Successes and failures of the \kp method: {A} direct assessment for
  {GaAs/AlAs} quantum structures.
\newblock {\em Phys. Rev B}, 53:7949--7963, 1996.

\bibitem{Yusa1998}
G.~Yusa, S.~J. Allen, J.~Davies, H.~Sakaki, H.~Kono, J.~Ahopelto, H.~Lipsanen,
  M.~Sopanen, and J.~Tulkki.
\newblock Teraherz-near infrared upconversion in strain-induced quanutum dots.
\newblock In {\em the 24th ICPS}, page 1083, 1998.

\end{thebibliography}

\clearpage
\newcolumntype{.}{D{.}{.}{-1}}
\newcolumntype{,}{D{,}{,}{2}}
\begin{table}
\caption{Material parameters from Landolt-B\"ornstein (Ref. \cite{LB}) unless otherwise noted.}
\label{tab:EC}
\begin{tabular}{>{ }l...}
\hline
\hline
Parameter
& \multicolumn{1}{c}{GaAs}
& \multicolumn{1}{c}{In$_{0.1}$Ga$_{0.9}$As}
& \multicolumn{1}{c}{InP} \\
\hline
$a_0$ (\AA)				&  5.65	&  5.69	&  5.87 \\
$c_{11}$ ($10^{11}$ dyn cm$^{-2}$)	& 11.90	& 11.50	& 10.11 \\
$c_{12}$ ($10^{11}$ dyn cm$^{-2}$)	&  5.38	&  5.29	&  5.61 \\
$c_{44}$ ($10^{11}$ dyn cm$^{-2}$)	&  5.95	&  5.75	&  4.56 \\
$e_{14}$ (C m$^{-2}$)   		& -0.16	& -0.15	& -0.04$ (Ref. \cite{Singh})$ \\
$\varepsilon_r$                         & 12.53	& 12.70	& 13.90 \\
\hline
\hline
\end{tabular}
\end{table}

\begin{table}
\caption{Eight-band ${\bf k} \cdot {\bf p}$ parameters from Ref. \cite{Vurgaftman2001}}
\label{tab:Mat}
\begin{tabular}{l..}
\hline
\hline
Parameter
& \multicolumn{1}{c}{GaAs}
& \multicolumn{1}{c}{In$_{0.1}$Ga$_{0.9}$As}\\
\hline
$E_{g}$ (eV)                           	&  1.519	& 1.366	\\
$\Delta$ (eV)                          	&  0.341	& 0.332	\\
$\gamma_1$                           	&  2.556	& 3.464	\\
$\gamma_2$                           	& -0.152	& 0.295	\\
$\gamma_3$                           	&  0.718	& 1.148	\\
$m_e$                                  	&  0.067	& 0.062	\\
$E_p$\footnote{These are the used value of $E_p$ and are
slightly smaller than the values given
in Ref.~\cite{Vurgaftman2001}. This was as explained in the
text.} (eV)
					& 20.160	&19.742	\\
$VBO$ (eV)               		& -0.055   	& 0.000	\\
$a_{v}$ (eV)                           	&  1.160	& 1.144	\\
$a_{c}$ (eV)                           	& -7.170	&-7.196	\\
$b_{v}$ (eV)                           	& -2.000	&-1.980	\\
$d_{v}$ (eV)                           	& -4.800	&-4.680	\\
$\varepsilon_r$\footnote{Ref. \cite{LB}}& 12.530	&12.727	\\
\hline
\hline
\end{tabular}
\end{table}
\clearpage
\begin{figure}
\begin{center}
\includegraphics[width=8.5cm,angle=0]{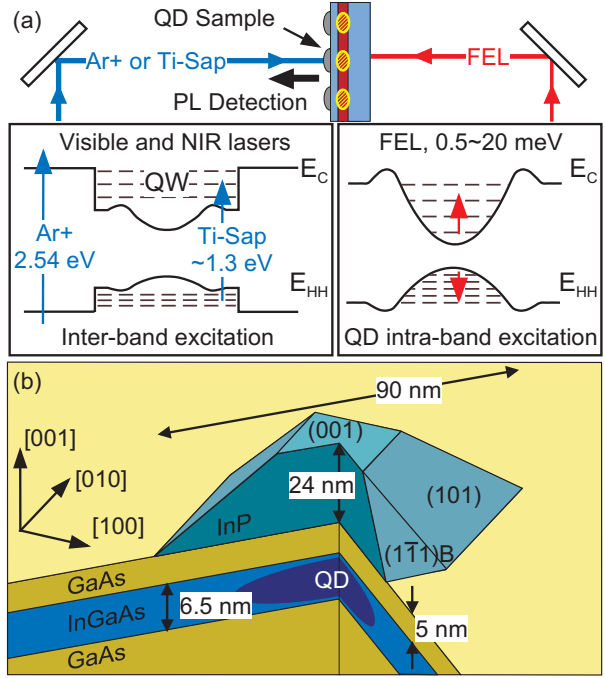}
\end{center}
\caption{
(Color online)
(a) In the analyzed experiment (Ref. \cite{Yusa1998}) the QD
sample was pumped by either an Ar+ or Ti-Sap laser. The QD
intraband dynamics was modulated by a free electron laser
(FEL). (b) Strain-induced quantum dot in an
In$_{0.2}$Ga$_{0.8}$As QW, strained by an InP island. The
island is defined by the crystal planes $\{001\}$,
$\{101\}$, and $\{111\}$.
}
\label{fig1}
\end{figure}
\clearpage
\begin{figure}
\begin{center}
\includegraphics[width=8.5cm,angle=0]{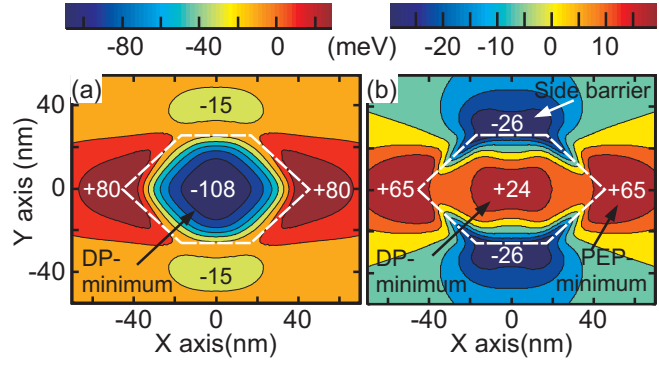}
\end{center}
\caption{
(Color online)
Potential energy of (a) electrons and (b) holes, in the
middle of the QW ($8.25~$nm below the InP island). The
energies are given with respect to the band edges of an
In$_{0.2}$Ga$_{0.8}$As QW, far away from the island. The
white dashed lines show the bottom contour of the InP
island and the labels correspond to the local potential
maxima/minima.
}
\label{fig2}
\end{figure}
\clearpage
\begin{figure}
\begin{center}
\includegraphics[width=8.5cm,angle=0]{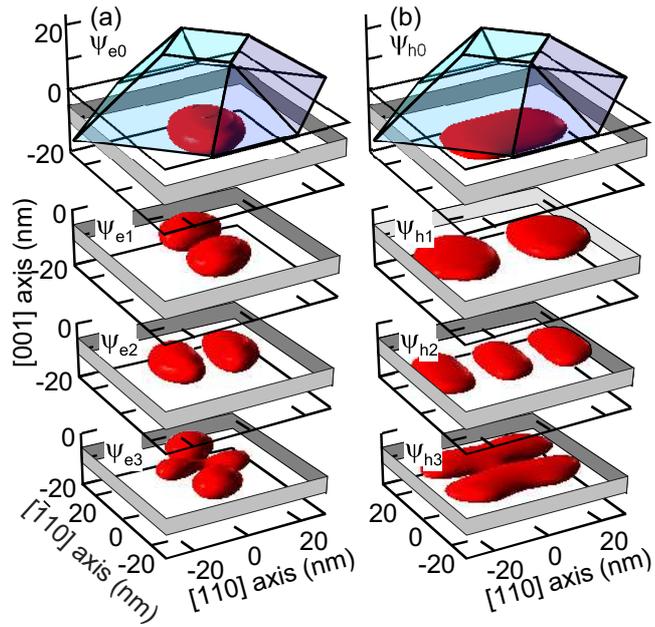}
\end{center}
\caption{
(Color online)
(a) Electron and (b) hole probability densities of the four
eigenstates, lowest in energy. The position of the QW is
shown by the gray faces.
}
\label{fig3}
\end{figure}
\clearpage
\begin{figure}
\begin{center}
\includegraphics[width=16.7cm,angle=0]{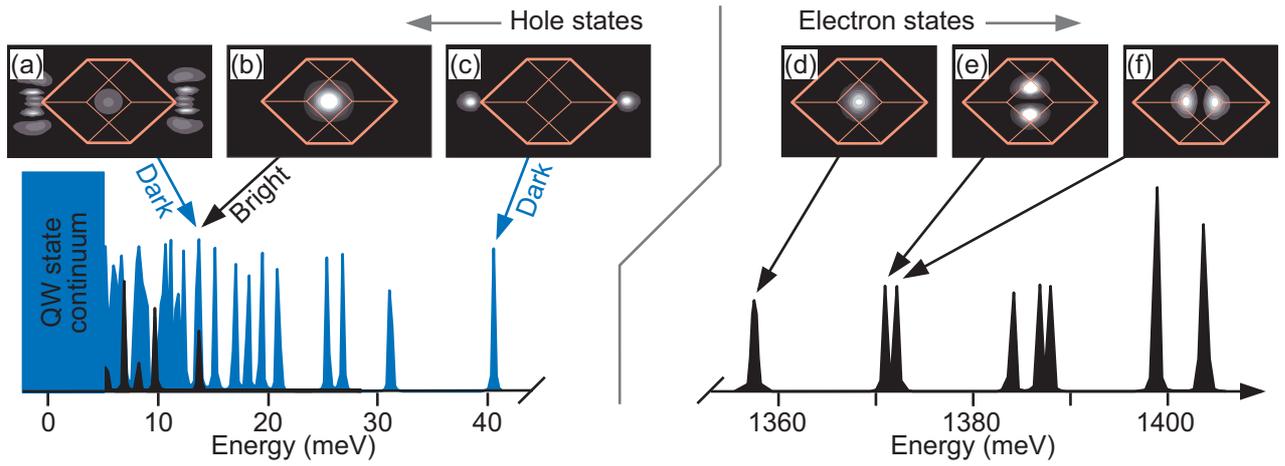}
\end{center}
\caption{
(Color online)
Density of hole and electron states, including states
confined in the DP and PEP minima. The blue/gray curve
is the total DOS while the black curve corresponds to
the DOS of carriers confined in the DP minima
only, i.e., the optically active states. The top
panels (a)-(f) show the probability densities of selected
states, with respect to the position of the InP island (red
lines) above the QD.
}
\label{fig4}
\end{figure}
\clearpage
\begin{figure}
\begin{center}
\includegraphics[width=8.5cm,angle=0]{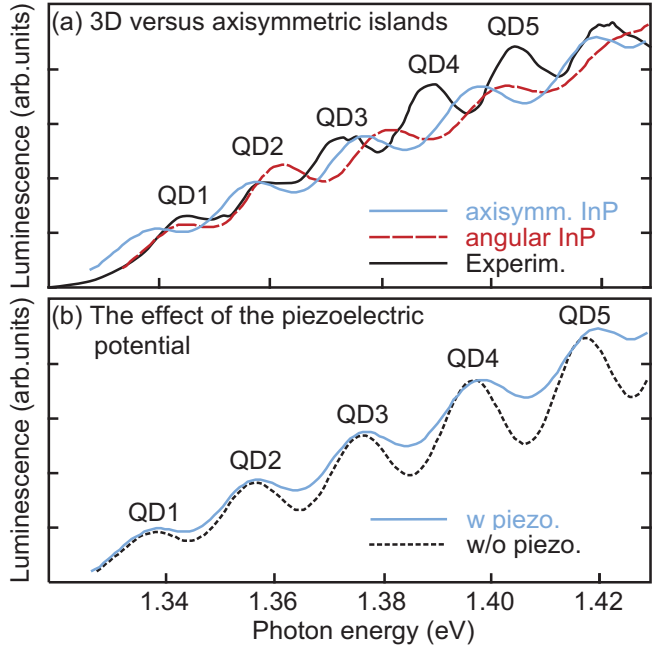}
\end{center}
\caption{
(Color online)
Simulated CW PL (gray/colored lines) in comparison with
experimental (Ref. \cite{Lipsanen1995}) data (black solid line).
Figure \ref{fig5}(a) shows the effect of the InP island
geometry. Both theoretical PL curves are based on full 3D
simulations, including the piezoelectric potential.
However, the solid blue/gray line was computed with InP
islands having the shape of a truncated cone
(axisymmetric), whereas the dashed line correspond to
simulations based on the InP island geometry of
Fig.~\ref{fig1}. Figure~\ref{fig5}(b) shows the effect of
the piezoelectric potential in the case of an axisymmetric
InP island.  The numerical PL curves have been normalized
by equating their QD4 peak to the corresponding QD4 peak of
the experiments
}
\label{fig5}
\end{figure}
\clearpage
\begin{figure}
\begin{center}
\includegraphics[width=8.5cm,angle=0]{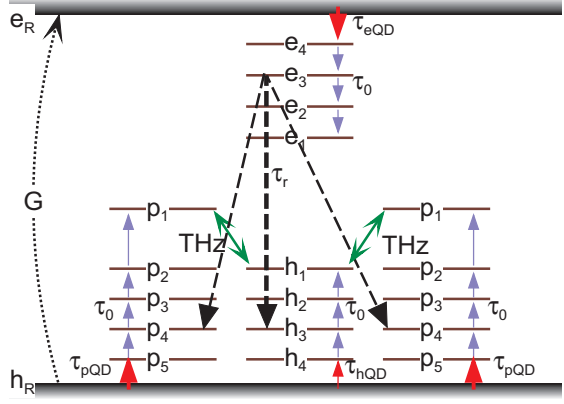}
\end{center}
\caption{
(Color online)
Carrier dynamics model. The eigenstates of the DP minima
are labeled $e_i$ and $h_i$; the hole states of the PEP
minima are labeled $p_i$. The pump laser $G$ generates
carriers to the reservoirs ($e_R$ and $h_R$) and the
resonant THz radiation transfers holes between the ground
states $p_1$ and $h_1$. Radiative recombinations (black
dashed arrows) are shown for the third electron level
only. The energy scale is only schematic.
}
\label{fig6}
\end{figure}
\clearpage
\begin{figure}
\begin{center}
\includegraphics[width=8.5cm,angle=0]{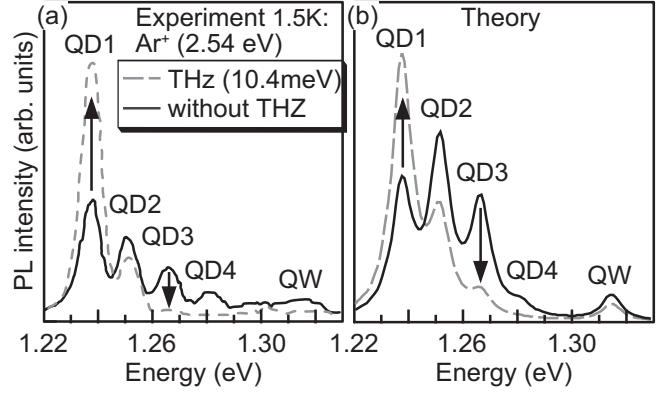}
\end{center}
\caption{
Experimental (a) and theoretical (b) photoluminescence
from a sample pumped by an Ar$^+$ laser only (solid lines)
and by Ar$^+$ laser and FEL simultaneously (dashed line).
The FEL energy of the experiments was
$\hbar\omega_{THz}=10.4~$meV.
}
\label{fig7}
\end{figure}
\clearpage
\begin{figure}
\begin{center}
\includegraphics[width=8.5cm,angle=0]{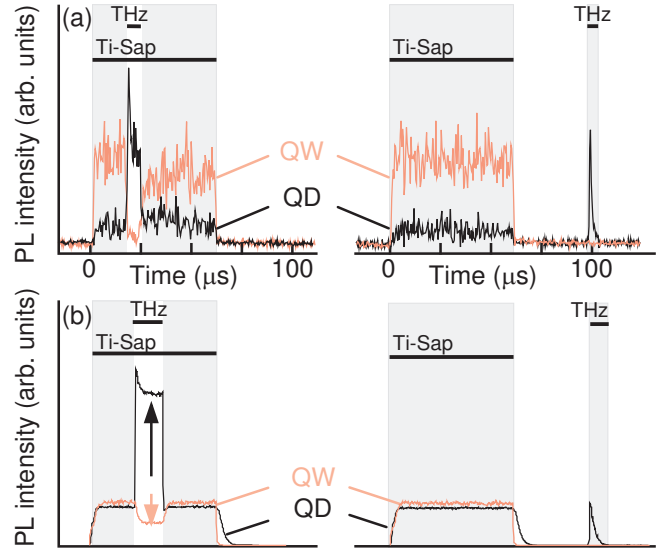}
\end{center}
\caption{
(Color online)
Experimental (a) and simulated (b) time-resolved PL of the
QD ground state (black) and QW (red) for $T=15~$K. The
energy and power of the THz radiation were
$\hbar\omega_{THz}=2.5~$meV and $P=1~$kW, respectively. The
time windows of the Ti-Sap laser (carrier generation to the
QW) and THz radiation are indicated by horizontal bars.
}
\label{fig8}
\end{figure}
\end{document}